\documentclass[prd,preprintnumbers,amsmath,amssymb]{revtex4}

\input epsf
\begin{document}
\draft

\title{Black hole quantum tunnelling and black hole entropy correction}
\author{Jingyi Zhang\footnote{E-mail: physicz@tom.com}} \affiliation{ Center for Astrophysics ,
Guangzhou University, 510006, Guangzhou, China}
\date{\today}

\begin{abstract}
Parikh-Wilczek tunnelling framework, which treats Hawking
radiation as a tunnelling process, is investigated again. As the
first order correction, the log-corrected entropy-area relation
naturally emerges in the tunnelling picture if we consider the
emission of a spherical shell. The second order correction of the
emission rate for the Schwarzschild black hole is calculated too.
In this level, the result is still in agreement with the unitary
theory, however, the entropy of the black hole will contain three
parts: the usual Bekenstein-Hawking entropy, the logarithmic term
and the inverse area term. In our results the coefficient of the
logarithmic term is $-1$. Apart from a
 coefficient, Our correction to the black hole entropy is consistent with that of loop quantum
 gravity.
\\\\PACS number(s): 04.70.Dy
\end{abstract}
\maketitle
\vskip2pc
\section{Introduction}
In 2000, Parikh and Wilczek proposed an approach to calculate the
emission rate at which particles tunnel across the event
horizon\cite{Parikh1}. They treat Hawking radiation as a
tunnelling process, and the WKB method is
used\cite{Parikh2,Parikh3}. In this way a corrected spectrum,
which is accurate to the first order approximation, is given.
Their result is considered to be in agreement with an underlying
unitary theory. Following this method, a lot of static or
stationary rotating black holes are studied
\cite{Hemming,Medved,Alves,Vagenas1,Vagenas2,Vagenas3,Vagenas4,Vagenas5,Vagenas6,Vagenas7,Vagenas8,Zhang1,Zhang2,Liu,Wu,Zhang3,Zhang4,Zhang5,Zhang6,Zhang7,Zhang8,Zhang9,Zhang10,Jiang,Majhi1,Majhi2,Majhi3,Kar}.
The same result, that is, Hawking radiation is no longer pure
thermal, unitary theory is satisfied and information is conserved,
is obtained. But in all of these literature, the entropy of the
black hole only contains the Bekenstein-Hawking entropy. Will the
emission process still consist with the unitary theory if the
quantum correction of the entropy is taken into account? In
present, about the quantum correction, there are different results
corresponding to different models and
methods\cite{Rovelli,Ashtekar,Kaul,Strominger,Solodukhin,Ghosh,Domagala,Meissner,Medved2,Kastrup,Gour,Chaterjee}.
The general formulation of the black hole entropy
is\cite{Medved4,Arzano}
\begin{equation}
S_q=\frac{A_H}{4l^2_p}+ \alpha
\ln{\frac{A_H}{4l^2_p}}+O(\frac{l^2_p}{A_H})+const.,\label{S}
\end{equation}
where $\alpha$ is a model-dependent (dimensionless) parameter. In
the case of Loop Quantum Gravity $\alpha$ is a negative
coefficient whose exact value was once an object of debate (see
e.g. \cite{Ghosh}) but has since been rigorously fixed at
$\alpha=-1/2$. In String Theory the sign of $\alpha$ depends on
the number of field species appearing in the low energy
approximation \cite{Solodukhin}. Therefore, a very interesting
work is to introduce the log-corrected entropy-area relation in
the tunnelling framework. Moreover, if the emission rate is
calculated to the second order approximation, will the entropy
contain the inverse area term as given in equation (\ref{S})? In
this paper, we first show that in the tunnelling picture, a
logarithm correction term occurs in the expression of the black
hole entropy if we take the emission particle as a spherical
surface wave (spherical shell). Therefore, we expect that
Parikh-Wilczek tunnelling framework, in fact, is in agreement with
the unitary theory to the first order log-correction of the black
hole entropy. Then, we verify that, if we calculate the emission
rate to the second order approximation with Parikh-Wilczek
tunnelling framework and suppose the emission process to be still
in agreement with the unitary theory, the entropy of the black
hole will contain three parts: the usual Bekenstein-Hawking
entropy, the logarithmic term and the inverse area term. Finally,
we give two comments to the Parikh-Wilczek framework and our
calculation.
\section{black hole tunnelling and the first order correction to the black hole entropy}
As mentioned above, Parikh and Wilczek applied the WKB
approximation to calculate the emission rate of a tunnelling
particle (S-shell). We start with a brief review of the WKB method
and the barrier penetration. For a massless particle (massless
shell), because of the infinite blueshift near the horizon, the
characteristic wavelength of any wavepacket of the S-wave (see
\cite{Parikh1,Parikh2,Parikh3}) is always arbitrarily small there,
so that the geometrical optics limit becomes an especially
reliable approximation. The geometrical optics limit allows us to
obtain rigorous results directly in the language of particles.
That is, the WKB method and the expression of the emission rate
are the same as that of a classical massive particle. So, in the
following discussion we only study the tunneling process of a
massive particle (massive shell).

 Schr$\ddot{\rm o}$dinger's equation for the motion of a particle in a centrally symmetric field is
\begin{equation}
\Delta \psi +(2m/(\hbar)^2)(E-U(r))\psi=0.\label{sfunction1}
\end{equation}
Let us consider the following radial equation:
\begin{equation}
\frac{1}{r^2}\frac{d}{dr}(r^2\frac{dR}{d
r})-\frac{l(l+1)}{r^2}R+\frac{2m}{\hbar^2}(E-U(r))R=0.\label{sfunction4}
\end{equation}
By the substitution
\begin{equation}
R(r)=X(r)/r \label{sfunction5}
\end{equation}
equation (\ref{sfunction4}) is brought to the form
\begin{equation}
\frac{d^2X}{dr^2}+[\frac{2m}{\hbar^2}(E-U(r))-\frac{l(l+1)}{r^2}]X=0.\label{sfunction6}
\end{equation}
For S-wave, $l=0$, the equation of $X(r)$ is
\begin{equation}
\frac{d^2X}{dr^2}+\frac{2m}{\hbar^2}(E-U(r))X=0.\label{sfunction8}
\end{equation}
Note that, in the Parikh-Wilczek framework, to calculate the
self-gravitation reliably the tunnelling particle is considered as
a spherical shell (S-wave). In this way, when it emits from the
black hole the matter-gravity system transits from one spherical
state to another. So, the de-Broglie wave function of the emission
spherical shell should be
\begin{equation}
\psi(r)=X(r)/r. \label{sfunction51}
\end{equation}
That is, the WKB wave function of a particle can be written as
\begin{equation}
\psi(r)=X
(r)/r=\frac{1}{r}\exp{[\frac{iS(r)}{\hbar}]},\label{wavefunction}
\end{equation}
where
\begin{equation}
S(r)=S_0(r)+(\frac{\hbar}{i})S_1(r)+(\frac{\hbar}{i})^2S_2(r)+\cdots.\label{phase}
\end{equation}
Substituting (\ref{wavefunction}) into Schr$\ddot{\rm o}$dinger
Equation (\ref{sfunction8}) yields
\begin{equation}
S_0=\pm\int^rp_r\,\mathrm{d}r,
\end{equation}
\begin{equation}
2S'_0S'_1+S''_0=0,
\end{equation}
\begin{equation}
2S'_0S'_2+(S'_1)^2+S''_1=0,
\end{equation}
where we use a prime to denote differentiation with respect to
$r$.

To evaluate the probability of a particle passing through the
barrier, we divide the whole region of motion of the particle by
two tunnelling points $a$ and $b$ into three parts: ingoing and
reflecting region I, barrier region I\!I and the outgoing region
I\!I\!I. The particle moves as a free particle in region I and
I\!I\!I, but region I\!I is classically inaccessible.

In region I, we take the WKB wave function as follows \cite{Zeng}
\begin{eqnarray}
X_{I}(r)&=&\frac{2}{\sqrt v}\sin{[\frac{1}{\hbar}\int^a_rp_r\,\mathrm{d}r+\frac{\pi}{4}]}\nonumber\\
&=&\frac{1}{i\sqrt{v}}\{\exp{[\frac{i}{\hbar}\int^a_rp_r\,\mathrm{d}r+\frac{i\pi}{4}]}-\exp{[-\frac{i}{\hbar}\int^a_rp_r\,\mathrm{d}r-\frac{i\pi}{4}]}\},
\end{eqnarray}
where $v$ is the velocity of the tunnelling particle. In region
I\!I, the WKB wave function is a linear combination of real
exponentials. Considering the connexion between the oscillating
and the exponential solutions at $r=a$, the WKB wave function in
region I\!I can be written as

\begin{equation}
X_{II}(r)=\frac{1}{\sqrt v}\exp{[-\frac{1}{\hbar}\mid\int^b_a
p_r\,\mathrm{d}r\mid]}\exp{[-\frac{1}{\hbar}\mid\int^r_b
p_r\,\mathrm{d}r\mid]}\label{connexion1}.
\end{equation}
And the WKB wave function in region I\!I\!I is
\begin{equation}
X_{III}(r)=-\frac{1}{\sqrt v}\exp{[-\frac{1}{\hbar}\mid\int^b_a
p_r\,\mathrm{d}r\mid]}\exp{[\frac{i}{\hbar}\int^r_b
p_r\,\mathrm{d}r+\frac{i\pi}{4}]}\label{connexion1}.
\end{equation}
The probability of barrier penetration is
\begin{equation}
\Gamma_p=\frac{j_{out}}{j_{in}}=\frac{v|\psi_{out}|^2}{v|\psi_{in}|^2}=\frac{v(X_{out}(b)/b)^2}{v(X_{in}(a)/a)^2}=\frac{a^2}{b^2}\cdot
\exp{[-\frac{\rm 2Im S_0}{\hbar}]}\label{Gamma}.
\end{equation}

Let's now calculate the phase space factor corresponding to the
black hole tunnelling. For Schwarzschild black hole, the line
element in Painlev$\acute{\rm e}$ coordinates is
\begin{equation}
\rm{d}s^2=-c^2(1-\frac{2MG}{c^2r})\rm{d}t^2+2c\sqrt{\frac{2MG}{c^2r}}\rm{d}t\rm{d}r+\rm{d}r^2+r^2(\rm{d}\theta^2+\sin^2{\theta}\rm{d}\phi^2),\label{line}
\end{equation}
and the radial null geodesics are
\begin{equation}
\dot{r}=\frac{\rm{d}r}{\rm{d}t}=\pm
c\,(1-\sqrt{\frac{2MG}{c^2r}}\,\,)\label{null}.
\end{equation}
with the upper(lower) sign in Eq. (\ref{null}) corresponding to
outgoing(ingoing) geodesics, under the implicit assumption that
$t$ increases towards the future\cite{Boyarsky}.

But in this paper we consider the tunneling of the massive
particle. That is, the outgoing particle is a massive shell (de
Broglie s-wave). The massive quanta doesn't follow
radial-lightlike geodesics (\ref{null}). Similar to Ref.
\cite{Zhang3}, we treat the massive particle as a de Broglie wave
and obtain the expression of $\overset{.}{r}$. Namely,
\begin{equation}
\overset{.}{r}=v_{p}=\frac{1}{2}v_{g}=-\frac{1}{2}\frac{g_{00}}{g_{01}}
=\frac{1}{2r}\frac{c^2r^{2}-2MGr}{\sqrt{2MGr}}. \label{vp}
\end{equation}
Note that to calculate the emission rate correctly, we should take
into account the self-gravitation of the tunnelling particle with
energy $\omega$. That is, we should replace $M$ with $M-\omega$ in
(\ref{line}) and (\ref{vp}) to describe the motion of the particle
correctly\cite{Parikh1,Parikh2,Parikh3}.

The canonical momentum $p_r$ and the imaginary part of the action
$\mathrm{Im}S_0$ can be easily obtained. Namely,
\begin{equation}
p_r=\int_0^{p_r}dp'_r=\int\frac {dH}{\dot{
r}}=-i\pi\frac{\hbar}{l^2_p}\,r,\label{p}
\end{equation}
\begin{equation}
\rm Im S_0=\rm \int_{r_i}^{r_f}p_r
dr=-\frac{1}{2}\hbar[\frac{A_f}{4l_p^2}-\frac{A_i}{4l_p^2}].\label{p}
\end{equation}
The probability of barrier penetration is
\begin{equation}
\Gamma_p=\frac{r^2_i}{r^2_f}\exp{[-\frac{\rm 2Im
S_0}{\hbar}]}=\exp{[(\frac{A_f}{4l_p^2}-\ln{\frac{A_f}{4l_p^2}})-(\frac{A_i}{4l_p^2}-\ln{\frac{A_i}{4l_p^2}})]},\label{ee}
\end{equation}
where $l_p^2=\frac{\hbar G}{c^3}$. In this paper, we investigate
the transition of the matter-gravity system from one spherical
state to another at the same energy. This transition corresponds
to the production and barrier penetration of the massive spherical
shell (or massless shell). That is, this process contains two
stages. The first stage is the production of the spherical shell
from the vacuum fluctuation near the event horizon. The second
stage is the barrier penetration. The rate of transition from the
initial spherical state to the final spherical state is therefore
\begin{equation}
\Gamma(i\to
f)=\Gamma_v\cdot\Gamma_p=\Gamma_v\cdot\exp{[(\frac{A_f}{4l_p^2}-\ln{\frac{A_f}{4l_p^2}})-(\frac{A_i}{4l_p^2}-\ln{\frac{A_i}{4l_p^2}})]}.\label{Gamma1}
\end{equation}
Let's Compare (\ref{Gamma1}) with the unitary result in Quantum
Mechanics, $\Gamma(i\to f)=\mid M_{fi}\mid ^2\cdot (\rm{phase\
space\ factor})$, which is given in Ref. \cite{Parikh1}. $\mid
M_{fi}\mid ^2$ is the probability amplitude of the process, in
this case it is related to the production rate of the particle in
the vacuum fluctuation near the event horizon. Thus, we obtain
\begin{equation}
\mathrm{phase\ space\
factor}=\exp{[(\frac{A_f}{4l_p^2}-\ln{\frac{A_f}{4l_p^2}})-(\frac{A_i}{4l_p^2}-\ln{\frac{A_i}{4l_p^2}})]}.\label{phasefactor2}
\end{equation}
If we bear in mind that
\begin{equation}
\mathrm{phase\ space\
factor}=\frac{N_f}{N_i}=\frac{e^{S_f}}{e^{S_i}}=e^{S_f-S_i},\label{phasefactor3}
\end{equation}
we naturally get the expression of the black hole entropy to the
first order correction
\begin{equation}
S_q=\frac{A_H}{4l_p^2}-\ln{\frac{A_H}{4l_p^2}}.\label{entropy}
\end{equation}

\section{second order correction to the black hole entropy}
Let's now calculate the tunnelling rate to the second order
approximation. In order to get the second order correction of the
black hole entropy, we write the WKB wave function to the second
order approximation. Namely,
\begin{equation}
X
(r)=\exp{[\frac{iS_0(r)}{\hbar}+S_1(r)+\frac{\hbar}{i}S_2(r)]},\label{wavefunction3}
\end{equation}
where
\begin{equation}
S_2=\int^r-\frac{(S_1^{'2}+S''_1)}{2S'_0}\mathrm{d}r.
\end{equation}

Like the treatment in section I\!I, the wave function in region I
can be taken as
\begin{eqnarray}
X_{I}(r)&=&\frac{2}{\sqrt v}\sin{[\frac{1}{\hbar}(\int^a_rp_r\,\mathrm{d}r-\hbar^2S_2(r))+\frac{\pi}{4}]}\nonumber\\
&=&\frac{1}{i\sqrt{v}}\{\exp{[\frac{i}{\hbar}(\int^a_rp_r\,\mathrm{d}r-\hbar^2S_2(r))+\frac{i\pi}{4}]}-\exp{[-\frac{i}{\hbar}(\int^a_rp_r\,\mathrm{d}r-\hbar^2S_2(r))-\frac{i\pi}{4}]}\}.
\end{eqnarray}
In this region the expression of $S_2(r)$ is
\begin{equation}
S_2=\int^a_r-\frac{(S_1^{'2}+S''_1)}{2S'_0}\mathrm{d}r.
\end{equation}
In order to reduce to the first order approximation case, the
connexion between the oscillating and the exponential solutions at
$r=a$ should be
\begin{equation}
\frac{2}{\sqrt
v}\sin{[\frac{1}{\hbar}(\int^a_rp_r\,\mathrm{d}r-\hbar^2S_2(r))+\frac{\pi}{4}]}\rightleftharpoons\frac{1}{\sqrt{
v}}\exp{[-\frac{1}{\hbar}(\int^r_a\mid
p_r\mid\,\mathrm{d}r-\hbar^2S_2(r))]}.\label{connexion3}
\end{equation}
\begin{equation}
r<a\qquad\qquad\qquad\qquad\qquad\qquad r>a\nonumber
\end{equation}
On the right hand of the connexion (\ref{connexion3}), the
expression of $S_2(r)$ is
\begin{equation}
S_2=\int^r_a-\frac{(S_1^{'2}+S''_1)}{2S'_0}\mathrm{d}r.
\end{equation}
The connexion at $r=b$ is
\begin{equation}
\frac{1}{\sqrt{ v}}\exp{[\frac{1}{\hbar}(\mid\int^r_b
p_r\,\mathrm{d}r\mid-\hbar^2S_2)]}\rightleftharpoons-\frac{1}{\sqrt{v}}\exp{[\frac{i}{\hbar}(\int^r_b
p_r\,\mathrm{d}r-\hbar^2S_2)+\frac{i\pi}{4}]},\label{connexion4}
\end{equation}
\begin{equation}
r<b\qquad\qquad\qquad\qquad\qquad r>b\nonumber
\end{equation}
and the wave function in region I\!I\!I is
\begin{equation}
X_{III}(r)=-\frac{1}{\sqrt{v}}\exp{[-\frac{1}{\hbar}(\rm Im
S_0-\hbar^2 Im S_2)}]\exp{[\frac{i}{\hbar}(\int^r_b
p_r\,\mathrm{d}r-\hbar^2S_2)+\frac{i\pi}{4}]},
\end{equation}
where
\begin{equation}
\mathrm{Im}S_2=\mathrm{Im}\int^b_a-\frac{(S_1^{'2}+S''_1)}{2S'_0}\mathrm{d}r.
\end{equation}
Since
\begin{equation}
\psi(r)=X(r)/r, \label{sfunction511}
\end{equation}
in region I, the ingoing flux density is
\begin{equation}
j_{in}=\frac{-i\hbar}{2m}(\psi_{in}\frac{\partial}{\partial
r}\psi_{in}^*-\psi_{in}^*\frac{\partial}{\partial
r}\psi_{in})=v|\psi^{2}_{in}|=\frac{1}{a^2}\label{ji},
\end{equation}
and in region I\!I\!I the outgoing flux density is
\begin{equation}
j_{out}=\frac{-i\hbar}{2m}(\psi_{out}\frac{\partial}{\partial
r}\psi_{out}^*-\psi_{out}^*\frac{\partial}{\partial
r}\psi_{out})=v|\psi^{2}_{out}|=\frac{1}{b^2}\exp{[-\frac{2}{\hbar}(\rm
Im S_0-\hbar^2 Im S_2)]}\label{jo}.
\end{equation}
Therefore,
\begin{equation}
\Gamma_p= j_{out}/j_{in}=\frac{a^2}{b^2}\exp{[-\frac{2}{\hbar}(\rm
Im S_0-\hbar^2 Im S_2)]}\label{tt}.
\end{equation}
For Schwarzschild black hole tunnelling, in classically
inaccessible region, we have
\begin{equation}
S'_0=p_r=-i\pi\frac{\hbar}{l^2_p}r,\quad
S''_0=-i\pi\frac{\hbar}{l^2_p}\label{sss1},
\end{equation}
and
\begin{equation}
S'_1=-\frac{1}{2}\frac{S''_0}{S'_0}=-\frac{1}{2r},\quad
S''_1=\frac{1}{2r^2}\label{ss}.
\end{equation}
From (\ref{ss}) we can easily obtain
\begin{equation}
S'_2=-\frac{1}{2S'_0}(S^{'2}_1+S''_1)=-(\frac{3i}{8\pi}\frac{l_p^2}{\hbar})\cdot\frac{1}{r^3}.
\end{equation}
So,
\begin{equation}
S_2=\int_{r_i}^{r_f}S'_2\,\mathrm{d}r=\frac{3i}{4\hbar}(\frac{l_p^2}{A_f}-\frac{l_p^2}{A_i})\label{s2}.
\end{equation}
Substituting (\ref{p}),(\ref{s2}) into (\ref{tt}) and considering
$\Gamma(i\to f)=\mid M_{fi}\mid ^2\cdot (\rm{phase\ space\
factor})$, yields
\begin{equation}
\mathrm{phase\ space\
factor}=\exp{[(\frac{A_f}{4l_p^2}-\ln{\frac{A_f}{4l_p^2}}+\frac{3}{2}\frac{l_p^2}{A_f})-(\frac{A_i}{4l_p^2}-\ln{\frac{A_i}{4l_p^2}}+\frac{3}{2}\frac{l_p^2}{A_i})]}.\label{phasefactor5}
\end{equation}
Comparing (\ref{phasefactor5}) with (\ref{phasefactor3}), we get
the expression of the black hole entropy to the second order
correction
\begin{equation}
 S_q=\frac{A_H}{4l_p^2}-\ln{\frac{A_H}{4l_p^2}}+\frac{3}{2}\frac{l_p^2}{A_H}+const.\label{sq1},
\end{equation}
which is consistent with an unitary theory and is also in
agreement with the general formulation of the black hole entropy.
The emission rate is
\begin{equation}
\Gamma(i\to f)\sim e^{\Delta S_q}\label{sq2}.
\end{equation}
\section{conclusion and comments}
We showed how a log-corrected entropy-area relation can emerge in
the tunnelling picture if we consider the emission particle as a
spherical shell. We also showed that, if the emission rate is
calculated to the second order approximation, the black hole
entropy will contain three parts: the usual Bekenstein-Hawking
entropy, the logarithmic term and the inverse area term. In our
calculation the logarithmic term and the inverse area term is the
consequence of requesting the process to satisfying the unitary
theory. Apart from a
 coefficient, Our correction to the black hole entropy is consistent with that of loop quantum
 gravity. In the following, we give two comments to the
 Parikh-Wilcek method and our calculation.

1) In this paper, we only take into account the emission of the
massive particle. The motion of a massless particle (S-wave) is
very different from that of a massive particle. However, as
mentioned in the first paragraph of the section II, a massless
shell can be treated in the language of particle. That is, for
massless shell we can also apply the WKB method and obtain the
same functional form of emission rate as that of the massive
particle. So, Eqs. (\ref{sq1}) and (\ref{sq2}) are also suitable
for massless particle's emission.

2) In the first order approximation, the previous expression of
the emission rate can be written in the following explicit form
\begin{equation}
\Gamma\sim \exp{(\Delta
S_q)}=(1-\frac{\omega}{M})^{\alpha}\exp{(-8\pi
GM\omega(1-\frac{\omega}{2M}))}.
\end{equation}
In Refs. \cite{Alves} and \cite{Vagenas7} the authors pointed out
that the coefficient of the log-corrected term in the black hole
entropy should be positive, otherwise, the probability of emission
will diverge when the emission particle's mass, $\omega$,
approaches to M. In fact, if we consider the applying condition of
the WKB method, the emission particle's mass, $\omega$, will never
approach to M, it is far smaller than the black hole mass M. Here
is our derivation.

The WKB method is established in the conditions:
\begin{equation}
\hbar|S^{''}_0|\ll|S^{'2}_0|\label{sss2},
\end{equation}
and
\begin{equation}
2\hbar|S^{'}_0 S^{'}_1|\ll|S^{'2}_0|\label{sss3}.
\end{equation}
From (\ref{sss1}) and (\ref{ss}), above conditions (\ref{sss2})
and (\ref{sss3}) can be incorporated into an inequality, that is
\begin{equation}
\hbar|\frac{dp_r}{dr}|\ll|p_r^2|\label{sss4}.
\end{equation}
Considering $p_r=-i\pi r$ and $2(M-\omega)\leqslant r\leqslant
2M$, (\ref{sss4}) becomes
\begin{equation}
2(M-\omega)\gg\sqrt{\frac{\hbar}{\pi}}\label{sss5}.
\end{equation}
That is,
\begin{equation}
M\gg\omega\label{sss6}.
\end{equation}
It means that the Parikh-Wilczek framework is only suitable for
the emission of the particle whose energy is far less than the
mass of the black hole. Most of the time in the evaporation of the
black hole this condition is satisfied, that is, the mass of the
emission particle will not approach to the black hole mass M.
Therefore, the coefficient of the log-corrected term in the black
hole entropy is not constrained to be positive. However, in the
last stage of evaporation the emission will be very strong, and
the mass of the emission particle will be very great, the WKB
conditions will not be satisfied, then one would have to resort to
other mechanisms to describe the last stage of the evaporation.
 \acknowledgments We thank Prof. S. Y. Pei and Prof. Z. Zhao for
helpful discussion. This research is supported partly by the
National Natural Science Foundation of China (Grant Nos. 10573005,
10633010),the National Basic Research Program of China (Grant No.
2007CB815405),and the Natural Science Foundation of Guangdong
Province (Grant No. 7301224).

\end{document}